\documentstyle[12pt,titlepage,fleqn]{article}

\newcommand{\be}{\begin{equation}}
\newcommand{\ee}{\end{equation}}
\newcommand{\ba}{\begin{eqnarray}}
\newcommand{\ea}{\end{eqnarray}}
\newcommand{\om}{\omega}
\newcommand{\q}{\hat{q}}
\newcommand{\p}{\hat{p}}

\newcommand{\non}{\nonumber}

\begin{document}

\begin{titlepage}
\begin{center}
{\large\bf  COLLECTIVE COORDINATES 
AND THE ABSENCE OF YUKAWA COUPLING IN THE CLASSICAL SKYRME MODEL }

\vskip 1.0cm

{\large F. ALDABE\footnote{E-mail: faldabe@phys.ualberta.ca}}

{\large\em Theoretical Physics Institute\\
University of Alberta\\
Edmonton, Alberta\\
 Canada, T6G 2J1}

\vspace{.25in}

\today \\

\vspace{.25in}

{\bf ABSTRACT}\\
\begin{quotation}
\noindent

In systems with constraints, physical states must
be annihilated by the constraints.  We make use of this rule to construct
physical
asymptotic states in the Skyrme model. The standard
derivation of the Born terms with asymptotic physical states shows that
there is no Yukawa
coupling for the Skyrmion.  
We propose a remedy tested in other
solitonic models: A Wilsonian action obtained after integrating the energetic
mesons and where
the Skyrmion is a quantum state should have a Yukawa coupling.

\end{quotation}
\end{center}

\end{titlepage}

The articles \cite{U,M,S}  have strongly claimed to have found 
a Yukawa coupling in the Skyrme model.  Although this claim is what
people want to hear, their results have been obtained in a way which is
inconsistent with systems with collective coordinates.  In \cite{M,S}, the
authors have failed
to take into account all the constraints.  In  \cite{U,M,S}, the authors
have failed to use asymptotic states which are physical.

It is very well known that the Skyrme model
has six constraints: three of them arise because the Skyrme solution
does not have the translational invariance of the model;
the other three are due to the rupture
of isospin invariance of the solution.  These
constraints may be written after linearization as \cite{U}
\ba
f^k&=&\int d^3x\sum_a -\frac{\partial\phi^a_S}{\partial x_k} \p^a\\ \label{cont}
f^a&=&\int d^3x\sum_{bc} i\epsilon_{abc} \phi_S^c \p^b \label{conr}
\ea
where $\p^a$ is the conjugate of $\q_a$. $\q_a$ and $\p_a$ 
are the fluctuations of the fields $\Phi^a$ and its conjugate
$P^a$ respectively,
which depend on the collective coordinates of the system (three translational,
$X^i(t)$, and three isospin coordinates, $\theta^a(t)$). $\phi^a_S$ is
the $a$th component of the Skyrme solution.

States which are physical must  be annihilated by the constraints \cite{bk}:
\be
f^n|phys>=0\label{con0}
\ee
where $n$ denotes any of the six constraints.
This has been the point which is missing in \cite{M,S,U} and which have
lead them to conclusions which are wrong.  
Consider a state with quantum number  $k$ and isospin $a$ which is expressed as
\cite{U1,O}
\be
|a\ k>=\int d^3x \psi_k(x,t) \stackrel{\leftrightarrow}{\partial}_o
q^a(x,t)|vac>,
\ee
and consider the case in which $k$ denotes momentum, and then
$\psi_k$ are plane waves solutions.
For $|a\ k>$ to be a physical state it must satisfy condition (\ref{con0}).
The vacuum must be physical which means 
\be
f^n|vac>=0.
\ee
for all $n$.  Then we find that 
\be
f^n|a\ k>=\int d^3x \psi_k(x,t) Y^n_a(\phi_s)(x)|vac>\ne0
\ee
where 
\be
Y^n_a(\phi_s)= i\epsilon_{anc} \phi_S^c 
\ee
for rotational constraints and 
\be
Y^n_a(\phi_s)= -\frac{\partial\phi^a_S}{\partial x_n} 
\ee
for translational constraints.
The
reason for the inequality follows from the fact that the 
$Y^n_a$'s, unlike the plane waves, are  not solutions
of the Klein-Gordon equations of motion.
Thus no orthogonality relation between $\psi_k$ and $Y^n_a$
can be established. 
This means that the states $|k,a>$ are not physical states and cannot
be used to construct asymptotic states.

The natural question is which states will be orthogonal to the $Y^n_a$.
The answer is simple: the use of Dirac brackets, $[,]_D$, which are
defined in terms of the Poisson brackets, $[,]_P$, the constraints and gauges,
must be used to obtain the equations of motion from the
Klein-Gordon Hamiltonian in  systems with 
collective coordinates \cite{T}.  The solutions to these equations of motion,
$f_r$, will then satisfy the twelve conditions
\be
\int d^3x f_r(x,t) Y^n_a(\phi_s)(x)|vac>=0.\label{e12}
\ee
Thus, states 
\be
|r\ a >=
\int d^3x f_r(x,t)\stackrel{\leftrightarrow}{\partial}_oq^a(x,t)|vac>,
\ee
 will be physical since they will satisfy condition
(\ref{con0}).
Because the $Y^n_a$ vanish at large distances from the soliton,
the  Dirac brackets tend to the Poisson brackets far away  from the soliton,
and thus the $f_r$'s will behave like plane waves 
for $|x|>>0$.

We must now construct the Born terms for the asymptotic states $|r\ a>$.
The Born terms can be obtained from the term in the reduction formula \cite{U}
\ba
&&-\int d^3x dt\int d^3x' dt'f^*_{r'}(x')f_{r}(x)\non\\
&&\; \; \; \; \; \; \; \; \; \; \; \; \; 
 <N'({\bf p}')|
T(-\nabla+m^2_{\pi})\phi_S^b(x')(-\nabla+m^2_{\pi})\phi_S^a(x)
|N({\bf p})>
\label{B1}
\ea
when the asymptotic states are taken to be plane waves. See \cite{U}
for notation.
When the solutions are not plane waves,
we must replace it by \cite{BD}
\be
-\int d^3x dt\int d^3x' dt'f^*_{r'}(x')f_{r}(x)
<N'({\bf p}')|
T\  \om_{r'}^2\phi_S^b(x')\om_r^2\phi_S^a(x) |N({\bf p})>
\label{B2}
\ee
since the step needed to arrive to (\ref{B1}) from (\ref{B2}) 
requires the use of the Klein-Gordon equation of motion which is not
the case at hand. 

It is a simple exercise following \cite{gjs, U} to arrive from (\ref{B2}) to 
\be
\frac{\om_r^2\int d^3x \nabla \phi_S^a f_r(x)\cdot
\int d^3x \nabla \phi_S^a f^*_r(x)}{M_S }+
\frac{\Delta_J\om_r^2 |\int d^3x \phi_S^a f_r(x)|^2 }{M_S }
\label{B4}
\ee
in the limit that the meson mass is smaller than the soliton mass.
Here $\Delta_J/M_S$ is the change in mass due to isospin rotations between
the state $|N(p)>$ and the intermediate baryonic state $|B(p')>$.
It is reassuringly that if the asymptotic states were plane waves, 
this expression is equivalent to eq. 3.25  of \cite{U} in the limit of 
large soliton mass.
In order to arrive to (\ref{B4}), 
the reader must avoid performing the integral
over space in (\ref{B2})  and make use of the fact that 
\ba
&&|p''-p|^2\int d^3x \phi_S^a e^{i (p''-p)\cdot x}
\int d^3x \phi_S^a e^{-i (p''-p)\cdot x}=\non\\
&&\;\;\;\;\;\;\;\;\;\;\;  \;\;\;\;\;\;\;\;\;\;  
         \int d^3x \nabla \phi_S^a e^{i (p''-p)\cdot x}\cdot 
\int d^3x \nabla \phi_S^a e^{-i (p''-p)\cdot x}
\ea

A close inspection of (\ref{B4}) shows that the asymptotic states
 must satisfy the conditions 
\ba
\int d^3x \nabla \phi_S^a f_r(x) &\ne&0\\
\int d^3x \phi_S^a f_r(x) &\ne&0
\ea
in order not to vanish. But this is equivalent to saying that the asymptotic
states do not satisfy the constraints (\ref{cont}) and (\ref{e12})
 and therefore are unphysical states.
Thus the leading order contribution to (\ref{B4})
vanishes, and the next to leading order
terms cannot be use to claim the existence of a Yukawa coupling.

We have therefore learned that there is no contribution to the Born terms to
leading order when the asymptotic states satisfy the constraints of the
system. Thus means that there is no Yukawa coupling for the Skyrmion as
claimed by other authors who have incorrectly used asymptotic states which
do not satisfy the constraint.

A possible objection is that it is not possible to replace the fluctuations by 
their classical fields in the reduction formula. However, as shown
in \cite{O},  the Born terms,  can also be obtained  by considering the
process
\be
\sum_{b m}\frac{|<r,a|V^{(2)}|m,b>|^2}{\om_m^2-\om_r^2}.\label{t1}
\ee
where $H^{(2)}=H^{(2)} _o+V^{(2)} $, $H^{(2)}$ is the total quadratic 
Hamiltonian,
$H_o^{(2)}$ is the Klein-Gordon Hamiltonian, and $V^{(2)}$ is the remaining
quadratic interactions.  Also, $\om_r$ are the eigefrequencies of the 
states $|r\ a>$ which diagonalize $H_o^{(2)}$and $\om_m$ are the 
eigenfrequencies  of the states $|m\ a>$ which diagonalize $H^{(2)}$.
Both type of states $|r\ a>$ and $|m\ a>$ diagonalize
their respective Hamiltonians by using the Dirac brackets and not the
Poisson brackets. A simple manipulation of (\ref{t1}) along the lines
of \cite{O} yields
\be
\sum_{n,b}\frac{|\int d^3x f_r Y^n_b|^2}{\om_r^2}+
\sum_{b,m}\ '\frac{|<r,a|V^{(2)}|m,b>|^2}{\om_m^2-\om_r^2}.\label{t2}
\ee
where the $\sum'$ is over positive energy states only.  The Born terms, is the
first term in (\ref{t2}).  Again we see that because the states $f_r$ are 
physical, the first term in (\ref{t2}) vanishes which means that there are
no Born terms.

This absence of Yukawa coupling for a classical Skyrmion should not be
taken as a drawback.  Rather, we should try to extend the work of \cite{YO2}
to show that the Wilsonian action obtained for a quantum Skyrmion after 
integrating out the very energetic mesons has a Yukawa coupling.

\pagebreak

\end{document}